\documentstyle[prb,preprint,aps]{revtex}

\begin{document}
\draft
\title{Effects of ac-field amplitude on the dielectric susceptibility of relaxors}
\author{Zhi-Rong Liu$^{*}$ and Bing-Lin Gu}
\address{Department of Physics, Tsinghua University, Beijing 100084, People's Republic of China}
\author{Xiao-Wen Zhang}
\address{Department of Materials Science and Engineering, 
Tsinghua University, Beijing 100084, People's Republic of China}
\maketitle

\begin{abstract}
The thermally activated flips of the local spontaneous polarization in
relaxors were simulated to investigate the effects of the applied-ac-field
amplitude on the dielectric susceptibility. It was observed that the
susceptibility increases with increasing the amplitude at low temperatures.
At high temperatures, the susceptibility experiences a plateau and then
drops. The maximum in the temperature dependence of susceptibility shifts to
lower temperatures when the amplitude increases. A similarity was found
between the effects of the amplitude and frequency on the susceptibility.
\end{abstract}


\vspace{2mm}



Relaxor ferroelectrics (relaxors) have been studied for nearly 40 years
since Pb(Mg$_{1/3}$Nb$_{2/3})$O$_3$ (PMN) was first synthesized by Smolenski 
{\it et al.}.\cite{1} The dielectric response of relaxors is characterized
by the diffuse phase transition (DPT) and a strong frequency dispersion.\cite
{2} Various models, such as the compositional heterogeneity model,\cite{1}
the superparaelectric model,\cite{2} and the glasslike model,\cite{3} {\it %
et al.}, were proposed to rationalize the complicated behaviors of relaxors.
It is widely accepted nowadays that the presence of polar
microregions in nanoscale\cite{4,5,6} is responsible for the relaxor
behaviors.

The effects of the applied ac field on relaxors\cite{7,8,9,10,11,12,13,14} 
cause great interest since
they provide some clue of the relaxation mechanism. Glazounov {\it et al.}
observed that the dielectric permittivity of PMN increases with increasing
amplitude of the applied ac field.\cite{7} A similarity was also found between
the effects of the amplitude and frequency on the permittivity.  In
addition, the ac-drive-enhanced relaxor characteristics and domain breakdown
were observed in (PbLa)(ZrTi) (PLZT).\cite{11} There are two possible
mechanisms, i. e., domain-wall motion model and superparaelectric model, to
explicate the nonlinearity of dielectric permitivity of PMN relaxors. 
Glazounov {\it et al.} suggested\cite{7} that it is related to domain-type
process rather than thermally activated flips of the local spontaneous
polarization (i.e. superparaelectric model). 
 However, they did not
consider the interaction of polar microregions when investigating the 
superparaelectric model, which is just one of the key
points related to response of the external field\cite{3,15,16}. In this study,
we conduct a Monte Carlo simulation to investigate the influence of
measuring field on the dielectric susceptibility of relaxors.


We investigate the thermally activated flipping process of the local
spontaneous polarization in relaxors. Following the work of Gui {\it et al.},
\cite{15} the polar microregions are regarded as point dipoles. Then relaxors
are modeled to be a system consisting of Ising-like dipoles with randomly
distributed interactions:\cite{15} 
\begin{equation}
H=-\sum\limits_{i\not{= }j}\stackrel{\sim}{J_{ij}}\sigma_{i}%
\sigma_{j}-E_{ext} \overline {\mu}\sum\limits_{i}{\frac{|\mu_{i}\cos%
\theta_{i}| }{\overline{\mu}}}\sigma_{i},
\end{equation}
where $\sigma_{i},\sigma_{j}=\pm1$ are dipole spins. When the projection 
of the $i$th dipole moment ${\vec{\mu }}_{i}$ on the direction of the 
external field ${\vec{E}}_{ext}$ is positive, $\sigma_i$ takes value +1, 
otherwise, $\sigma_i$ takes value -1.
$\theta_{i}$ is
the angle between ${\vec{\mu }}_{i}$ and ${\vec{E}}_{ext}$, and 
$\overline{\mu}$ is the maximal magnitude of the dipole moments. 
$\stackrel{\sim}{J_{ij}}$ is the effective interaction energy
between the nearest neighbor dipoles, which has a Gaussian distribution with 
a width $\Delta J$. $\stackrel{\sim}{J_{ij}}$
reflects the correlation between polar microregions, which is essential to
the glassy behaviors.\cite{3,15,16} In general, the external field contains
a measuring ac field and a bias dc field. In this paper, only the ac field
is involved, {\ i.e.}, 
\begin{equation}
E_{ext}=E_0\exp\left(i2\pi\frac{t}{t_L}\right),
\end{equation}
where $t$ is the real time. $E_0$ and $t_L$ are the amplitude and the period
of the ac field, respectively.

The Monte Carlo simulation is performed on a $16\times 16\times 16$ simple
cubic lattice with periodic boundary conditions. The details 
of simulation process can be found in Ref. 15.
The dielectric susceptibility is defined as
\begin{equation}
\chi=C\left \langle \frac {\frac{1}{t_{obs}}\int\limits_{t_0}^{t_0+t_{obs}}
p(t)\exp\left(i2\pi\frac{t}{t_L}\right)dt } {E_{ext}} \right\rangle,
\end{equation}
where $C$ is a proportional factor which is chosen to be 1 in this
contribution, and $\langle\cdots\rangle$ denotes the configurational
averaging. $p(t)$ is the normalized polarization:
\begin{equation}
p(t)=\frac{1}{N}\sum\limits_{i}{\frac{|\mu_{i}\cos\theta_{i}| }{\overline{\mu}}}
\sigma_{i}.
\end{equation}
During the simulation process, $p(t)$ is recorded and $\chi$ is calculated 
according to Eq. (3). 
We choose $t_0=200$MCS/dipole to eliminate the influence of the
initial state and $t_{obs}=3000$MCS/dipole to be the observation time.
The simulation is performed in many runs with different initial conditions 
so that the configurational averaging can be done. Longer observation time 
was also adopted in test, but no obvious influence on results was observed.


In order to verify the validity of the method, the dielectric susceptibility
under a weak field is firstly calculated. The result is shown in Fig. 1. It
can be seen that the susceptibility $\chi$ reaches its maximum at
a certain temperature ($T_m$) and changes gradually around $T_m$, which is
known as the diffuse phase transition (DPT) in relaxors. A strong frequency
dispersion can be also observed: $\chi$ decreases with increasing field
frequency at low temperatures, and $T_m$ moves to higher temperatures. All
these characteristics are consistent with the experiments\cite{2} and the
previous theoretical results\cite{15}.

Now, let us investigate the effects of the field amplitude on the dielectric
susceptibility. The susceptibility curves under different ac-field
amplitudes $E_0$ are depicted in Fig. 2 when the measuring frequency is kept
as $t_L=10$MCS/dipole. (We  express the frequency by $t_L$ here and
hereafter.) From Fig. 2 one can list the most essential features of the
nonlinear effect: (1) the dielectric susceptibility increases with
increasing $E_0$ at temperatures $T<T_m$ where the frequency dispersion is
observed; (2) increasing $E_0$ will make the maximum in the temperature
dependence of $\chi$ shifts to lower temperatures, which has the similar
effect of decreasing frequency (see also Fig. 1). 
The change of the imaginary part, $\chi''$, shows similar features in the 
simulation. These features agree with
the experiments in PMN\cite{7,8} very well. The concepts of ``slow dipole" 
and ``fast dipole" can
help to understand the increasing of the susceptibility. Slow dipoles are
those dipoles which flip too slow to keep up with the changing of the
ac field and give no or little contribution to the
dielectric susceptibility. At low temperatures, there are large amounts of
slow dipoles.\cite{15} When $E_0$ increases, the
driving force on slow dipoles is enhanced. Slow dipoles are forced to flip
faster and they give more contribution to the dielectric susceptibility 
$\chi$. For fast dipoles, the contribution changes slightly at low drives 
(see below). As a result, the susceptibility $\chi$ increases with increasing $E_0$.

It can be seen in Fig. 2 that the dielectric susceptibility slightly
decreases with increasing the external-field amplitude $E_0$ at high
temperatures. The tendency is weakened at higher frequencies while 
becomes more evident at lower frequencies.
Fig. 3 shows the cases for a lower frequency $t_L=50$MCS/dipole. It shows
that the dielectric susceptibility decreases at high temperatures and
increases at low temperatures when $E_0$ increases. These results are
similar to the experimental cases in PLZT\cite{11,12} to some extent. 
However, the computed maximum in $\chi(T)$ decreases with increasing $E_0$, 
which is opposite to the experimental observations.\cite{11} It 
reflects the defect of the model or/and the method we used.

To get further knowledge of the continuous effects of the ac-field
amplitude, we plot in Fig. 4 the curves of $\chi$ as functions of the
amplitude $E_0$ for different temperatures when the measuring frequency is
fixed as $t_L=10$MCS/dipole. At low temperatures, the dielectric
susceptibility increases first, and then drops with increasing $E_0$. This
means that the applied field speeds up the flipping of dipoles at small $E_0$
values so $\chi$ increases first, while the system is nearly saturated at
large $E_0$ values which causes the drop of $\chi$. At high temperatures,
the dielectric susceptibility experiences a plateau at the beginning and
then decreases when the applied field increases. These results are consistent 
with the experiments in PMN when $E_0$ varied in wide range of values.\cite{14} 
In Ref. 7 and Ref. 8, $E_0$ is not large enough, so 
$\chi$ increases at low temperatures and remains steady at high temperatures
with increasing $E_0$.

Fig. 5 demonstrates the field dependence of $\chi$ at different measuring
frequencies and a fixed temperature $T=1.5\Delta J/k_B$. It shows that the
maximum of the curve shifts to lower field amplitude when decreasing the
measuring frequency. The shapes of curves are similar for different
frequencies.

Fig. 6 shows the temperature of the susceptibility maximum ($T_m$)
as a function of the external-ac-field amplitude $E_0$. A nonlinear relation
can be found between $T_m$ and $E_0$. It is conflict with the linear law
observed in experiments\cite{7,8}. Perhaps the field used in experiments is
not large enough to reveal the high-order effects of the $T_m\sim E_0$
curve. Further experiments are needed to testify the theoretical predictions.

The Eq. (3) could be generalized to include the Fourier component at 
different frequencies than that of $E_{ext}$. Fig. 7 gives the curve 
of $\chi_{2\omega}/\chi$, where $\chi_{2\omega}$ is the second-order 
component of the susceptibility. It can be seen that $\chi_{2\omega}/\chi$ 
is stronger at lower $E_{ext}$ and $T$.

By means of the results above, we can see that the behaviors of the system
described by the model Hamiltonian in Eq. (1) are consistent with many 
aspects of the experiments when the applied-ac-field amplitude
varies. There are two points that should be mentioned here. First, the
interactions between polar microregions play an important role in the
dielectric response. If the interaction does not exist, the dielectric
susceptibility will decrease with increasing field amplitude as what is
pointed out by Glazounov {\it et al.}\cite{7,8}. Secondly, the model in Eq.
(1) is a rather simplified model. It cannot reflect the effects of external
field on the crystal structure completely.\cite{11} It describes only the
thermally activated flips process of the local polarization. Indeed, there
may be more dielectric mechanism in relaxors. For example, It was 
presented that there may be two kinds of polarization processes in 
relaxors.\cite{17} Very recently, various types of contributions were found 
to dominate the dielectric response within different ac-drive amplitude 
ranges.\cite{13}


In conclusion, the simulation results suggest that the thermally activated 
flips of the local
spontaneous polarization in relaxors plays an important role in producing
the relaxation phenomena.


This work was supported by the Chinese National Science Foundation.

\vspace{2mm}

* Electronic address: zrliu@phys.tsinghua.edu.cn

\begin{figure}[tbp]
\caption{Weak-field susceptibility as a function of temperature (in units of 
$\Delta J/k_B$). The field amplitude is fixed as $E_0$=0.1 
$\Delta J/\bar{\mu}$. The curves 1-4 correspond to the field
frequency $t_L=100,50,20,10$MCS/dipole, respectively. }
\end{figure}

\begin{figure}[tbp]
\caption{Dielectric susceptibility at various field amplitudes, $E_0$
(1-0.5, 2-1.0, 3-1.5, 4-2.0$\Delta J/\bar{\mu}$). The field frequency is
kept as $t_L=10$MCS/dipole. Inserted graphics is the imaginary part of 
susceptibility.}
\end{figure}

\begin{figure}[tbp]
\caption{Dielectric susceptibility at various field amplitudes, $E_0$
(1-0.5, 2-1.0, 3-1.5, 4-2.0$\Delta J/\bar{\mu}$). The field frequency is
kept as $t_L=50$MCS/dipole. }
\end{figure}

\begin{figure}[tbp]
\caption{Field amplitude (in units of $\Delta J/\bar{\mu}$) 
dependence of dielectric susceptibility at 
a fixed field frequency $t_L=10$MCS/dipole. 
Curves 1-5 correspond to temperatures 
T=0.5,1.5,2.5,3.5 and 4.5$\Delta J/k_B$, respectively. }
\end{figure}

\begin{figure}[tbp]
\caption{Field amplitude dependence of susceptibility at various frequencies
and a fixed temperature $T=1.5\Delta J/k_B$. The amplitude and the frequency
are measured in units of $\Delta J/\bar{\mu}$ and MCS/dipole, respectively.}
\end{figure}

\begin{figure}[tbp]
\caption{Temperature $T_m$, corresponding to the position of the maximum in $%
\chi(T)$, as a function of the field amplitude (in units of $\Delta J/\bar{%
\mu}$). The temperature is measured in units of $\Delta J/k_B$. }
\end{figure}

\begin{figure}[tbp]
\caption{Temperature dependence of $\chi_{2\omega}/\chi$ at various field 
amplitudes, $E_0$(1-0.5,2-1.5$\Delta J/\bar{\mu}$). The field frequency is 
kept as $t_L$=10MCS/dipole. }
\end{figure}

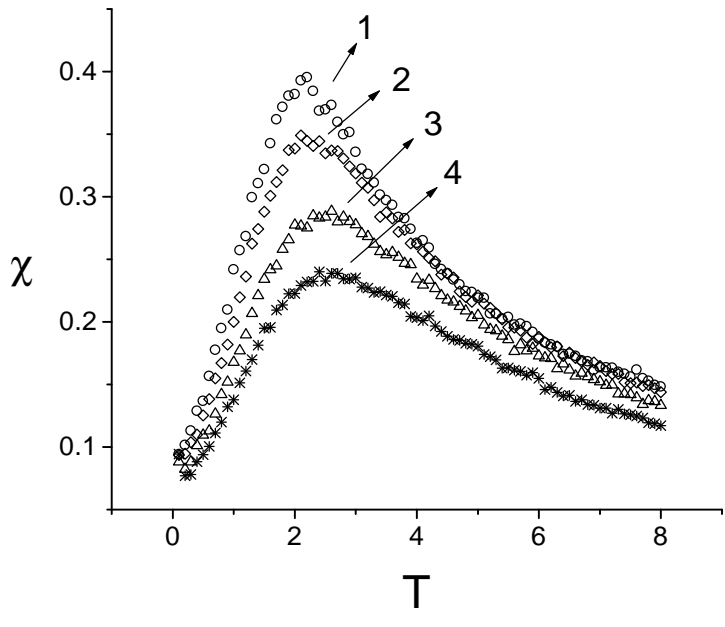
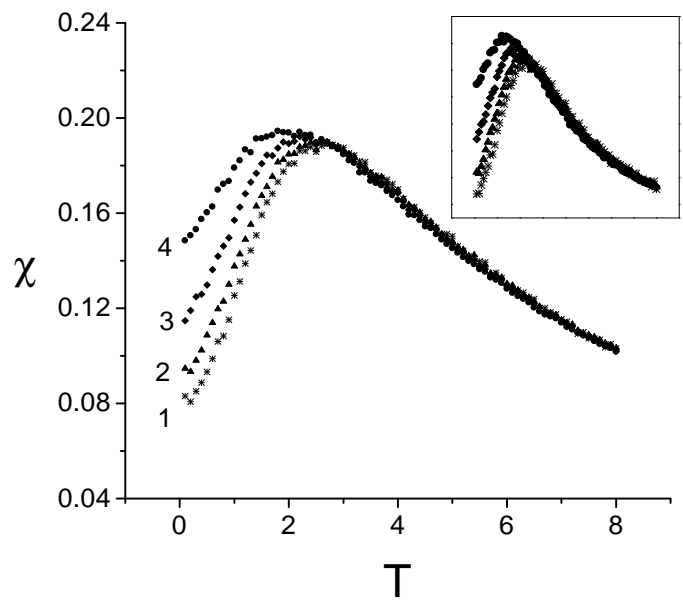
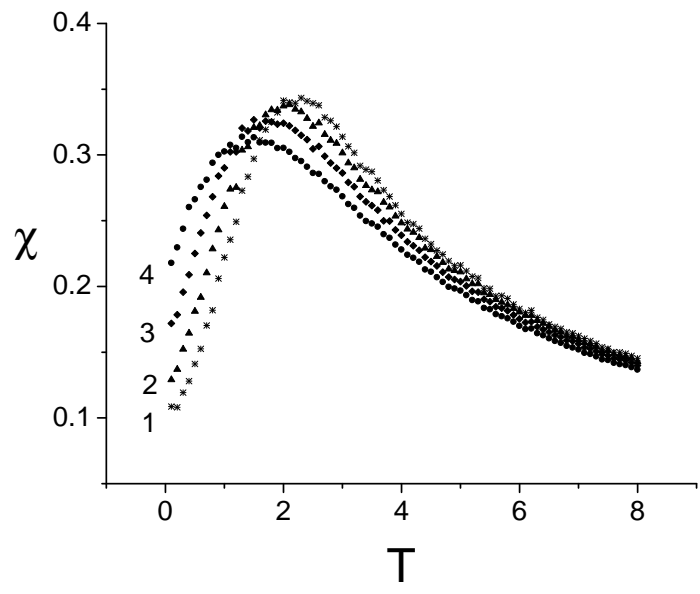
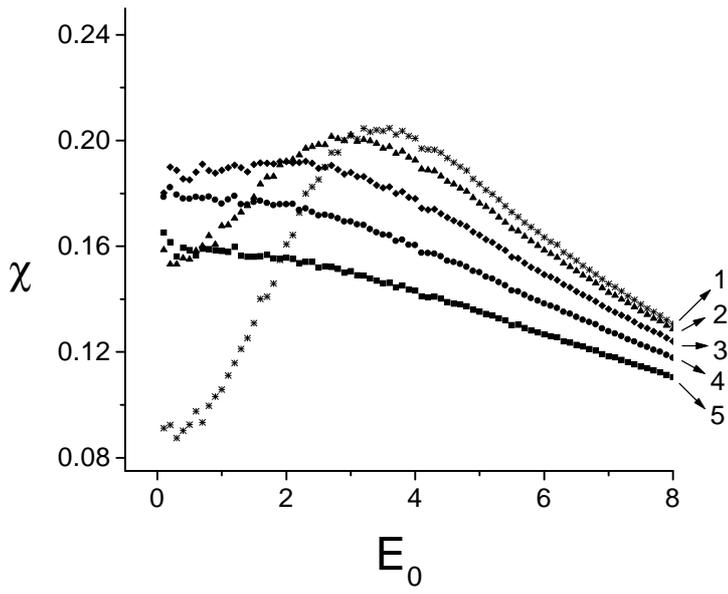
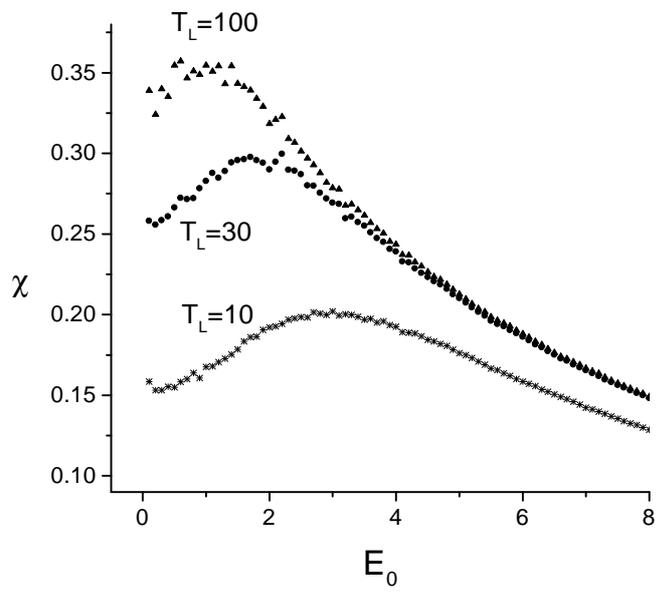
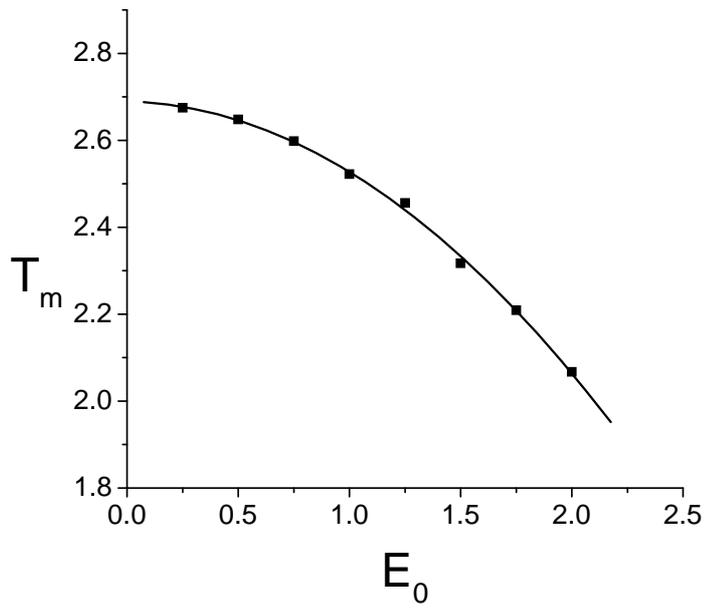
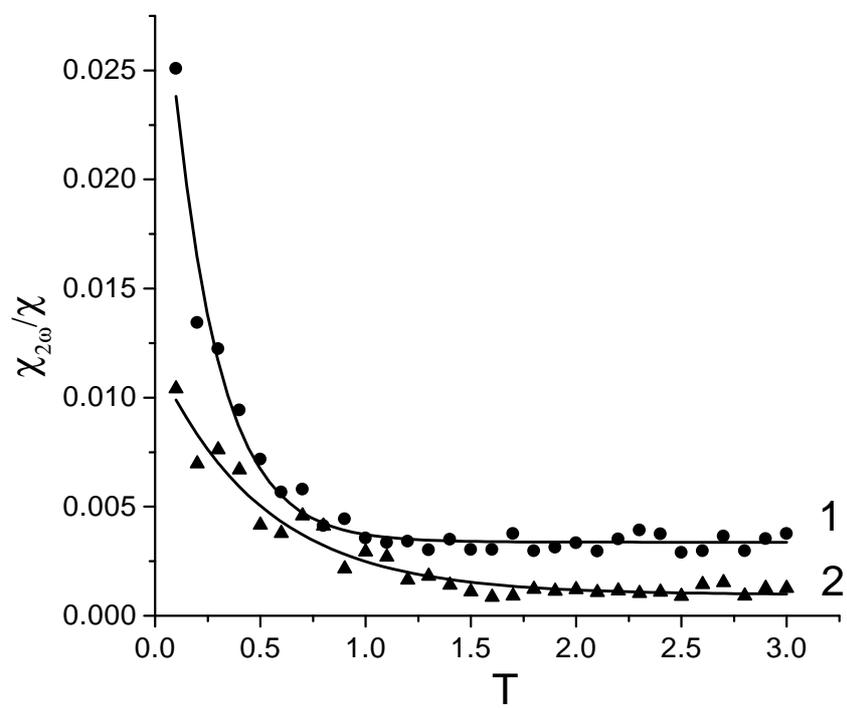
\end{document}